# Strategic Development of a Hydrogen Supply Chain in Corsica: a Multi-criteria Analysis

Tchougoune Moustapha Mai*
UMR CNRS 6134 Scientific Centre
Georges Peri
University of Corsica
Ajaccio, France
moustapha-mai_m@univ-corse.fr

Mohamed Hajajji
UMR CNRS 6134 Scientific Centre
Georges Peri
University of Corsica
Ajaccio, France
hajjaji_m@univ-corse.fr

Catherine Azzaro-Pantel
Laboratoire de Génie Chimique
Université de Toulouse
Toulouse, France
Catherine.azzopantel@toulouse-inp.fr

Maude Chin Choi
UMR CNRS 6134 Scientific Centre
Georges Peri
University of Corsica
Ajaccio, France
chin-choi_m@univ-corse.fr

Christian Cristofari
UMR CNRS 6134 Scientific Centre
Georges Peri
University of Corsica
Ajaccio, France
cristofari@univ-corse.fr

*Abstract*— A multi-objective framework for hydrogen supply chain (HSC) planning is developed for island contexts, incorporating Mixed-Integer Linear Programming (MILP) over multiple time periods. The model minimizes total system cost, greenhouse gas (GHG) emissions, and a risk index criteria. The case study of Corsica is considered, using Geographic Information Systems (GIS) for spatial analysis and infrastructure locating. The 2050 future design of the HSC is determined including site selection, capacity sizing, and technology choices. The proposed m-TOPSIS-based multi objectives solution shows a decentralized infrastructure with a levelized cost of hydrogen of €6.55/kg, and greenhouse gas emissions under 2 kgCO$_2$e/kg H$_2$. The study also integrates water availability and tourism-induced demand variation as key drivers of energy planning in insular regions.

## I. Introduction

According to IEA [1], the growth in global electricity generation capacity from solar panels, wind turbines and other renewable technologies will accelerate in the coming years. The massive integration of electricity produced from intermittent renewable energy sources will further strain the power grid. In the energy transition context, hydrogen is often considered as a clean and flexible alternative energy carrier that can satisfy a variety of applications (industry, heating, transport, power generation) [2]. According to some authors [3], Islands can stand out as valuable contributors to global development by becoming ideal platforms for testing and advancing sustainable hydrogen technologies.

Corsica, a Mediterranean Non-Interconnected Areas (NIA), possesses substantial potential for renewable energy, particularly from solar and wind sources. However, the island still heavily depends on imported fossil fuels to meet its energy needs. To address this challenge, Corsica has developed a roadmap that aims to achieve energy self-sufficiency by 2050. Unlike mainland France, where electricity generation is mainly dependent on nuclear power (66.6%), Corsica's energy mix is dominated by fossil fuels (40%). Geographically closer to Italy than France, Corsica is partially connected to the Italian grid through two links (mainland Italy and Sardinia), providing up to 26% of the island's electricity in 2022, as shown in Figure 1 [13,14].

The mobility sector is the largest source of GHG emissions in Corsica, accounting for more than 50% of the island's final energy consumption. As it is the most polluting sector and the primary target of the territorial decarbonization plan, our study focusses on this sector, with the aim of achieving hydrogen production of up to 8 tons per day by 2028 [6]. Following this, Corsica also plans to use hydrogen to decarbonize port and airport operations, as well as energy storage systems that support the electrical grid.

This study aims to outline pathways for deploying hydrogen systems in NIA, with Corsica as case study. The proposed methodology combines MILP with multi-criteria analysis, enabling a thorough assessment of system costs, GHG emissions, and supply chain relative risks. The developed framework can ensure both technical stability and economic sustainability of the future HSC. Beyond Corsica, the goal is to develop a generic model adaptable to other NIA, taking into account demographic patterns, water resources, and geographical conditions. The findings from this work could offer practical guidance for policymakers interested in improving energy autonomy on island territories.

This paper is organized as follows. Section 2 offers a state-of-the-art analysis of hydrogen deployment projects in international island territories. Section 3 reviews modeling studies focused on establishing the HSC specifically for islands. Section 4 outlines the methodological framework, detailing various techno-economic, environmental and risk factor parameters, along with associated mathematical constraints. Section 5 presents and discusses the island case study. Finally, Section 6 concludes the paper and suggests potential areas for improvement.

## II. Method and Tools

### A. General framework

The methodological framework for designing and operating a future HSC is based on a geographical and multi-objective approach aimed at meeting the hydrogen demand profile of a given region. The MILP problem is developed using the GAMS® modeling system and solved with the CPLEX 12 solver (De Leon Almaraz, 2014). Three main criteria are optimized: minimizing the global network cost (capital and operating expenditures), minimizing the GHG emissions and minimizing of risk index criteria [8] subject to: supply, demand, mass conservation and technical performance.

Input data of the model include various techno-economic parameters, such as production, storage, and transport units, available sites, energy resources, as well as the capital and operating costs of facilities, and the greenhouse gas emissions of different technologies. The model incorporates several constraints using continuous, integer, and binary variables. Integer variables define facility locations, system sizing, and the choice of appropriate production networks based on water and energy resource availability. Binary variables indicate the direction of hydrogen transport. To address the multi-criteria dimension, the ε-constraint method was employed, optimizing three objectives to build a Pareto front: cost minimization was treated as the main objective function, while environmental impact and safety were integrated as inequality constraints.

The best trade-off solution of the multi-objective optimization was obtained by calculation of the shortest distance between Utopia point and all combinations in the Pareto front (TOPSIS method, De-Leon Almaraz et al., [9].

A post-optimal step is then carried out for the obtained compromise solutions which are then evaluated to ensure their technical feasibility with the GIS tool.

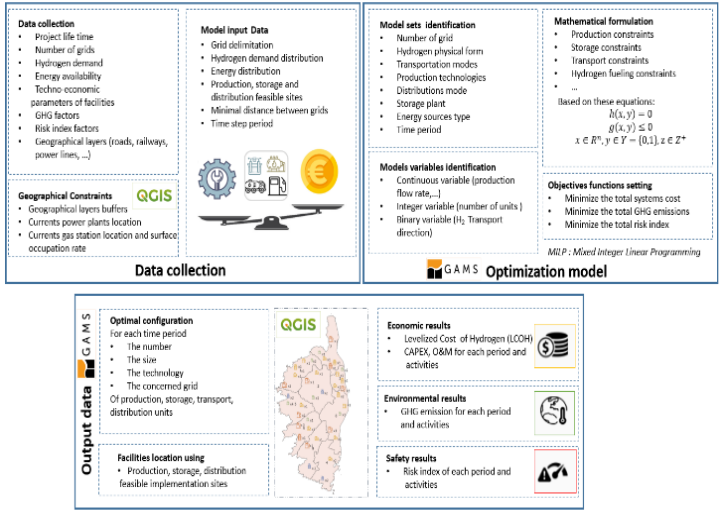

Fig. 1. Methodological framework overview

## B. Mathematical formulation

The novelty of this work is based on the monthly time period interval which allows to consider seasonal Island variations. Additional parameters such as the number of operating hours, the power of the electrolyzers, water availability constraints are also taken into consideration.

- **Energy supply constraints:** The total availability of primary energy sources in a given zone $g$ during period $t$ et and for month $m$ ($Atot_{gtm}$) is defined as the sum of three components: the initial average availability of renewable energy sources in each zone and for each month $A0_{gtme}$, the importation of primary energy sources from the grid $IPES_{egtm}$ and the consumption rate of primary energy sources $ESU_{gtm}$.

$ESU_{gtm}$ is defined as the product of $\gamma_{pj}$ which represents the energy consumption rate of production facities, and $PR_{pjigtm}$, the daily hydrogen production rate for each type of production facility $p$ and size $j$.

$$Atot_{g,t,m} = \sum_e (A0_{g,t,m,e} + IPES_{e,g,t,m}) - er \times ESU_{g,t,m}; \quad \forall\, e,g,t,g,m; g \neq g'; \quad (1)$$

$$ESU_{g,t,m} = \sum_{p,j,i}(\gamma_{p,j} \times PR_{p,j,i,g,t,m}); \forall g,t,m; g \neq g' \quad (2)$$

- **Energy source availability:** The total availability of photovoltaic and wind energy sources ($PV_{g,m,e}$ et $Wind_{g,m,e}$) is determined by considering the power capacity of the energy source in each network during period t and for each energy type ($ESP_{g,t,e}$), the number of hours per month $Mh_m$, the capacity factor for each energy source ($PVCF_m$ and $WindCF_m$, respectively for photovoltaics and wind energy), and the total number of days per month $Nd_m$.

$$PV_{g,m,e} = \frac{ESP_{g,m,e} \times Mh_m \times PVCF_m}{Nd_m} \quad \forall\, g,m,e; \quad g \neq g' \quad (3)$$

$$Wind_{g,m,e} = \frac{ESP_{g,m,e} \times Mh_m \times WindCF_m}{Nd_m} \quad \forall\, g,m,e; \quad g \neq g' \quad (4)$$

- **Water resource constraint:** Equations (5) and (6) describe the intermediate vulnerability value $waterVul_g$ and the final vulnerability value $waterVulsn_{g,m}$ of each grid. $SRFWaterVul_g$ and $SOTWaterVul_g$ represent the vulnerability index for surface water and groundwater, respectively. $VulSaison_{gm}$ represente the precipitation risk index during saison and grid distribution.

$$WaterVul_g = 0{,}8 \times SRFWaterVul_g + 0{,}2 \times SOTWaterVul_g; \forall\, g \quad (5)$$

$$WaterVulsn_{g,m} = WaterVul_g \times VulSaison_{g,m}; \forall\, g,m \quad (6)$$

The index of the quantity of water consumed by the final water vulnerability index $WCV_{g,t,m}$ is limited by lower and upper limits for the water vulnerability index ($WaterVul^{min}_{g,m}$; $WaterVul^{max}_{g,m}$) and water withdrawal rates ($minCW$ et $maxCW$). The total amount of potable water distributed in Corsica ($CleanWater_t$) is approximately 50 million m³ between 2012 and 2018 [10].

$$WaterVul^{min}_{g,m} \times CleanWater_t \times minCW \times 1000000 \leq WCV_{g,t,m} \leq WaterVul^{max}_{g,m} \times CleanWater_t \times maxCW \times 1000000; \forall\, g,t,m \quad (7)$$

- **Cost objective function:** the objective cost function is used to minimize the total cost of the system. This function is translated by the LCOH, which takes into account the system's periodic investment $CAPEX_t$, the $OPEX_t$, which reflects the system's maintenance and operating costs, and the quantity of hydrogen produced $DTtot_t$. $n3_{t,m}$ and $n4_{t,m}$ represent the period time index according to the year and the month.

$$LCOH = \sum_{m,t} \frac{\left[\frac{CAPEX_t}{[(1+f)\times(1+dr)]^{n3_{t,m}}}\right] + \left[\frac{OPEX_t \times Nd_m}{[(1+f)\times(1+dr)]^{n4_{t,m}}}\right]}{\frac{DTtot_t}{[(1+f)\times(1+dr)]^{n4_{t,m}}}}; \forall\, t \quad (8)$$

- *Global warning potential objective function:* The total global warming potential (*GWPtot*) is determined from the total daily greenhouse gas (*GHG*) emissions of the production $PGWP_t$, storage $SGWS_t$ and transport units $TGWT_t$.

$$GWPtot = \sum_t PGWP_t + SGWS_t + TGWT_t \quad (9)$$

- *Risk index objective function:* the total risk index of the HSC (*RiskTotal*) is determined considering the risk index of production $TPRisk_t$, storage $TSRisk_t$ and transport $TTRisk_t$ as described in [8].

$$RiskTotal = \sum_t TPRisk_t + TSRisk_t + TTRisk_t \quad (10)$$

## III. CORSICA STUDY CASE

### A. Problem description

This long-term optimization model for hydrogen deployment in Corsica is based on a territorial division into nine (9) strategic grids. Each grid is characterized by specific hydrogen demand and renewable energy potential. The model integrates spatial, technological, and temporal constraints, considering monthly variations in hydrogen production and delivery up to the year 2050. Two electrolyser technologies: Proton Exchange Membrane (PEM) and Alkaline (AE) are evaluated for hydrogen production, which is geographically restricted based on a preliminary GIS analysis. Storage units are allowed in all grids, ensuring a minimum three-day autonomy, while hydrogen transport is enabled via compressed or liquefied form using tanker-trucks and tube trailers. Only photovoltaic and wind energy sources are considered.

The model simultaneously optimizes the design and operation of production, storage, transport, and distribution units, while estimating key performance indicators such as cost, greenhouse gas (GHG) emissions, risk levels, and the levelised cost of hydrogen (LCOH). A multi-objective optimization approach is implemented using the ε-constraint method, followed by a multi-criteria decision analysis (m-TOPSIS) to support trade-off decisions among cost, environmental impact, and safety. Equal weighting is applied to each objective, ensuring balanced sustainability outcomes.

### B. Data collection

*Hydrogen Demand:* Corsica, a French island in the Mediterranean strongly shaped by seasonal tourism, experiences major fluctuations in energy demand over the course of the year. According to estimates from the Regional Climate, Air, and Energy Plan (SRCAE), hydrogen could account for around 2.5% of the island's total fuel consumption by 2030 [11]. This would correspond to a hydrogen demand varying between 4.3 tonnes per day in January and 8.1 tonnes per day in August. Figure 2 shows the monthly changes in hydrogen demand expressed in energy terms.

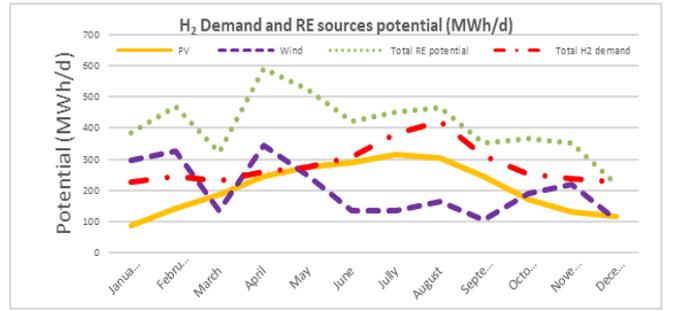

Fig. 2. H₂ Demand and RE source potential

*Techno-economic parameters:* Two technologies of electrolyzer are used in this study: Alkaline (AE) and Proton Exchange Membrane (PEM) electrolyzers. In Corsica, these technologies have been identified as the most suitable for the hydrogen supply chain, thanks to their technological maturity and their low dependence on high-temperature heat sources. The renewable energy capacities dedicated to hydrogen production (60 MW for photovoltaics and 40 MW for wind) were sized based on the island's available potential. Taking into account their respective capacity factors, Figure 2 provides an estimate of their projected output for the year 2030. Additionally, Table 1 lists the minimum and maximum values of key parameters related to production, storage, transport, distribution facilities, and the cost of electricity.

TABLE I. PARAMETERS OF THE TECHNO-ECONOMIC MODEL

| | Capacity | CAPEX | OPEX | Efficiency |
|---|---|---|---|---|
| **Production (Electrolyzer)** | 300-5 000 (kW) | 1038-3500 (€/kW) | 0.1-0.250 (€/kg) | 37.8-52 (kWh/kg) |
| **Conversion (Compressor)** | 126 (kg/h) | 13.41 €/(kg/h) | 0.007 (€/kg) | 2.66 (kWh/kg) |
| **Conversion (Liquefier)** | 50 (kg/h) | 149.2 €/(kg/h) | - | 6.78 (kWh/kg) |
| **Storage (pressurized tank)** | 50-30 000 (kg) | 25-500 (€/kg) | 0.006-0.02 (€/kg) | - |
| **Distribution (Refueling station)** | 20-1,300 (kg/d) | 410-1,480 (k€) | 0.15-0.39 (€/kg) | 52.4-56.4 (kWh/kg) |
| **Transport (Truck)** | 670-4,300 (kg) | 746-200 (€/kg) | Depending on the distance travelled | - |
| **Electricity price from energy sources** | | | | |
| | *ADEME Hypothesis 1 (A1) Reference trajectory* [12] | | *ADEME Hypothesis 2 (A2)* **Nuclear extension** [12] | |
| **Photovoltaics** | From €0.014/kWh to €0.019/kWh between 2025 and 2050 | | From €0.031/kWh (2025) to €0.034/kWh (2050) | |
| **Wind** | From €0.029/kWh to €0.044/kWh between 2025 and 2050 | | €0.031/kWh and €0.040/kWh between 2025 and 2050, with a low of €0.026/kWh in 2035 | |

## IV. RESULTS AND DISCUSSIONS

### A. Optimal hydrogen supply chain

The proposed multi-objective optimization model simulates a long-term deployment of HSC under multiple constraints and sustainability criteria. Using a Dell laptop equipped with a Core i5 processor and 16 GB of RAM, the model reaches convergence after approximately 17 hours of computation. The optimal configuration (presented in figure 3 for January 2050 HSC configuration) resulting from the multi-objective hydrogen supply chain model reveals a decentralized system structured around nine medium-sized storage units, each located in each grid. This setup ensures efficient coverage while optimizing investment and operating costs, emissions, and safety risks.

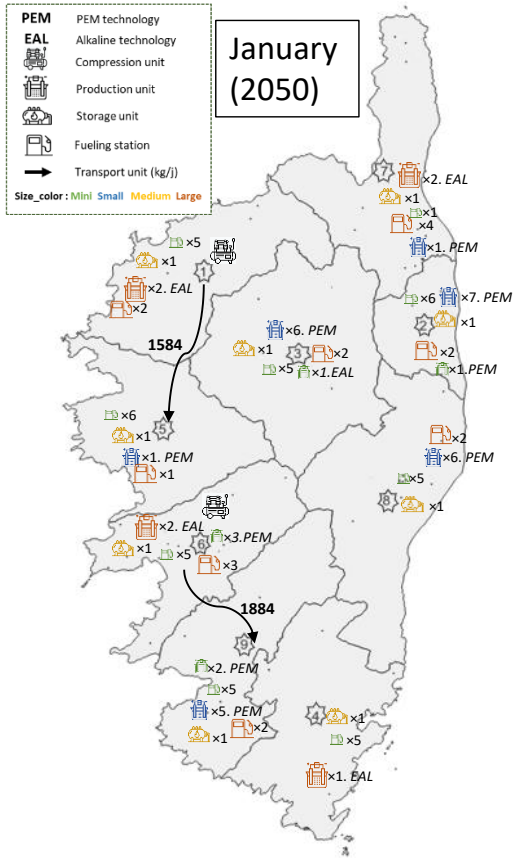

Fig. 3. Optimal hydrogen supply chain configuration for january 2050

By 2050, the system includes 43 hydrogen production units, predominantly Proton Exchange Membrane (PEM) type, mostly of mini and small size complemented by a smaller share of large-scale Alkaline units located in zones with high renewable energy potential. The deployment of these units is uneven over time, with several periods of no new installations.

The distribution subsystem shows 57 hydrogen fueling stations, mostly under 50 kg per day of capacity and some large-scale fueling station (1300 kg/day). Adjustments in station size and location are made to match demand across the different grids.

The number of transport unit remains stable, with key hydrogen flows between specific zones. A third transport units can become necessary toward the end of the period (2050) to meet growing demand in certain areas.

Overall, daily hydrogen exports flows across the HSC increase progressively from around 1,280 kg/day in 2026 to over 3,460 kg/day in 2050, with clear seasonal fluctuations higher in summer period.

*B. Optimal criterias Pareto font*

The developed model relies on the ε-constraint method to generate a diverse set of optimal solutions, each evaluated across three main criteria: total system cost (in k€/day), greenhouse gas (GHG) emissions (in $tCO_2e$/day), and a global risk index. These solutions are represented in Figure 4 through Pareto front: one linking cost and risk, the other linking cost and GHG emissions.

The decision-making process is supported by the m-TOPSIS method, which identifies a compromise solution using equal weights across all criteria. This baseline solution, illustrated by red triangles in Figure 4, corresponds to a cumulative system cost (Capex + Opex) of €1.55 million/day over a 25-year horizon (i.e., an annual average of 62.2 k€/day), GHG emissions of 549 $tCO_2e$/day (21.9 $tCO_2e$/day on average), and a global risk index of 1,325 (average index value of 53).

To further explore environmentally favorable scenarios, additional solutions were generated by modifying the weighting of the GHG emissions criterion in the m-TOPSIS analysis. When the weight of GHG emissions is doubled (two times more important), a solution with emissions reduced to 18 $tCO_2e$/day is obtained. Similarly, by excluding the risk index criterion, optimal solutions with emissions as low as 19 $tCO_2e$/day are identified depending on the selected weight for GHG.

Finally, incorporating constraints related to water accessibility across grids, shows minimal impact on the optimization results. The cost and risk-oriented solutions see a slight decrease in cost (about 0.7%), while GHG emissions remain unchanged. These constraints primarily affect the spatial distribution of hydrogen production units within the system, rather than its overall performance. Some grids with limited resources in global water availability are restricted to produce more hydrogen (medium and large-scale production unit are not installed).

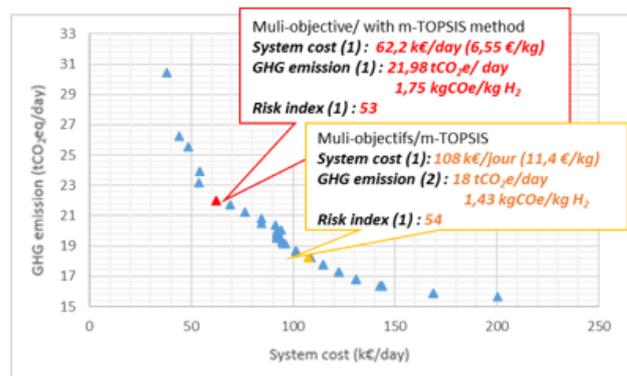

Fig. 4. Multi objective and bi-objective optimization Pareto Front

V. CONCLUSION

This research develops a methodological framework for planning and operation of a HSC in Corsica, explicitly incorporating monthly fluctuations in hydrogen demand, the availability of renewable energy resources, and water constraints. Beyond its technical dimension, this framework also provides a foundation for evaluating potential policy incentives aimed at fostering a hydrogen-based economy in NIA.

The multi-objective optimization of the hydrogen supply chain in Corsica outlines decentralized system projected to include 43 production units, 57 distribution stations, and 9 storage units by 2050. The system meets a growing hydrogen demand that reaches 2,202 kg/day by the end of the period, with seasonal peaks around 900 kg/day in summer. The optimized configuration achieves a balanced trade-off, with an average daily system cost of 62.2 k€, GHG emissions of 21.9 tCO$_2$e/day, and a global risk index of 53. These results provide a strategic framework for sustainable hydrogen deployment in island territories.

Further research should investigate advanced multi-objective optimization methodologies capable of capturing inter-seasonal dynamics in renewable energy availability and freshwater resources. In parallel, the development of integrated hydrogen water nexus models, specifically adapted to the spatial and infrastructural constraints of insular systems, could provide innovative strategies for the resilient and sustainable deployment of hydrogen technologies in regions such as Corsica.

ACKNOWLEDGMENT

This work has received research funding from the French government managed by the National Research Agency under the France 2030 program, with reference number ANR-22-EXES-016.

REFERENCE

[1] "Renewable electricity growth is accelerating faster than ever worldwide, supporting the emergence of the new global energy economy - News," IEA. Accessed: Mar. 08, 2022. [Online]. Available: https://www.iea.org/news/renewable-electricity-growth-is-accelerating-faster-than-ever-worldwide-supporting-the-emergence-of-the-new-global-energy-economy

[2] M. Kim and J. Kim, "An integrated decision support model for design and operation of a wind-based hydrogen supply system," *International Journal of Hydrogen Energy*, vol. 42, no. 7, pp. 3899–3915, Feb. 2017, doi: 10.1016/j.ijhydene.2016.10.129.

[3] G. Krajačić, R. Martins, A. Busuttil, N. Duić, and M. da Graça Carvalho, "Hydrogen as an energy vector in the islands' energy supply," *International Journal of Hydrogen Energy*, vol. 33, no. 4, pp. 1091–1103, Feb. 2008, doi: 10.1016/j.ijhydene.2007.12.025.

[4] "Information sur l'origine de l'électricité fournie par EDF | EDF FR." Accessed: Sep. 11, 2024. [Online]. Available: https://www.edf.fr/origine-de-l-electricite-fournie-par-edf

[5] EDF SEI, "Bilan électrique EDF sei 2017," 2018. Accessed: May 03, 2021. [Online]. Available: https://www.edf.com/

[6] M. BACCI, "La révision de la Programmation Pluriannuelle de l'Energie est lancée !," Agence d'aménagement durable, d'urbanisme et d'énergie de la Corse. Accessed: Apr. 06, 2022. [Online]. Available: https://www.aue.corsica/La-revision-de-la-Programmation-Pluriannuelle-de-l-Energie-est-lancee-_a377.html

[7] Sofia De Lèon Almaraz, "Multi-objective optimisation of a hydrogen supply chain," Institut National Polytechnique de Toulouse (INP Toulouse), INP Toulouse, 2014.

[8] J. Kim, Y. Lee, and I. Moon, "Optimization of a hydrogen supply chain under demand uncertainty," *International journal of hydrogen energy*, vol. 33, no. 18, pp. 4715–4729, Jan. 2008.

[9] S. De-León Almaraz, C. Azzaro-Pantel, L. Montastruc, and S. Domenech, "Hydrogen supply chain optimization for deployment scenarios in the Midi-Pyrénées region, France," *International Journal of Hydrogen Energy*, vol. 39, no. 23, pp. 11831–11845, Aug. 2014, doi: 10.1016/j.ijhydene.2014.05.165.

[10] Comité de bassin de Corse, "Tableau de bord du SDAGE," 2019. [Online]. Available: https://www.corse.eaufrance.fr/gestion-de-leau/dce-sdage/tableaux-de-bord-du-sdage-et-bilan

[11] "Le Schéma Régional Climat, Air, Energie (SRCAE) de Corse," Agence d'aménagement durable, d'urbanisme et d'énergie de la Corse. Accessed: Oct. 26, 2021. [Online]. Available: https://www.aue.corsica/Le-Schema-Regional-Climat-Air-Energie-SRCAE-de-Corse_a31.html

[12] "Trajectoires d'évolution du mix électrique à horizon 2020-2060," La librairie ADEME. Accessed: Oct. 27, 2021. [Online]. Available: https://librairie.ademe.fr/energies-renouvelables-reseaux-et-stockage/1173-trajectoires-d-evolution-du-mix-electrique-a-horizon-2020-2060-9791029711732.html